\documentclass{article}

\usepackage{bm}
\usepackage{dsfont}
\usepackage{amsmath}
\usepackage{amssymb}
\usepackage{graphicx}
\usepackage{xcolor}
\usepackage[mathcal]{euscript}
\usepackage{authblk}
\usepackage[colorlinks=true,linkcolor=blue,allcolors=blue]{hyperref}
\usepackage{geometry}
\newcommand\blfootnote[1]{%
  \begingroup
  \renewcommand\thefootnote{}\footnote{#1}%
  \addtocounter{footnote}{-1}%
  \endgroup
}

\title{\textbf{A Neural Network Architecture to Learn \\ Explicit MPC Controllers from Data}\footnote{This work has received support from the Swiss National Science Foundation under the RISK project (Risk Aware Data-Driven Demand Response, grant number 200021 175627.}}

\author{\normalsize E. T. Maddalena, C. G. da S. Moraes, G. Waltrich and C. N. Jones}
\date{}
\begin{document}
\maketitle
\thispagestyle{empty}
\vspace{-0.5cm}
\begin{abstract}
We present a methodology to learn explicit Model Predictive Control (eMPC) laws from sample data points with tunable complexity. The learning process is cast in a special Neural Network setting where the coefficients of two linear layers and a parametric quadratic program (pQP) implicit layer are optimized to fit the training data. Thanks to this formulation, powerful tools from the machine learning community can be exploited to speed up the off-line computations through high parallelization. The final controller can be deployed via low-complexity eMPC and the resulting closed-loop system can be certified for stability using existing tools available in the literature. A numerical example on the voltage-current regulation of a multicell DC-DC converter is provided, where the storage and on-line computational demands of the initial controller are drastically reduced with negligible performance impact.
\blfootnote{E. T. Maddalena and C. N. Jones are with \'Ecole Polytechnique F\'ed\'erale de Lausanne (EPFL), Switzerland (e-mails: \texttt{emilio.maddalena@epfl.ch}, \texttt{colin.jones@epfl.ch}). C. G. da S. Moraes and G. Waltrich are with Universidade Federal de Santa Catarina (UFSC), Brazil (e-mails: \texttt{caio.moraes@posgrad.ufsc.com}, \texttt{gierri.waltrich@ufsc.br}).} \\[10pt]
\end{abstract} \smallbreak

\section{Introduction}

Model Predictive Control (MPC) is a valuable control methodology to be considered whenever performance needs to be maximized, possibly operating the system close to its physical constraints. For certain classes of problems -- e.g. when the dynamics and constraints are affine and the objective is quadratic -- a closed-form solution exists, mapping the current state to the control inputs through a continuous piecewise-affine (PWA) function. This approach is known as explicit Model Predictive Control (eMPC) and is particularly employed when the hardware where such a control law is to be implemented cannot solve the MPC mathematical program on-line (either for time limitations or even safety regulations regarding the final application). Unfortunately, eMPC controllers might also suffer from having high complexity due to the exponential growth of the number of regions in the worst case \cite{alessio2009survey}, thus leading to not only high memory requirements but also high computational times. 

Scaling down the memory footprint and computational burden of eMPC is possible through a variety of techniques. Given a measurement of the current state, deciding in which region of the polyhedral partition the system is can be efficiently accomplished using search trees, hash tables, among other data structures \cite{jones2006logarithmic,bayat2011using,monnigmann2011fast}. If some information is known about the initial condition of the plant (even if only with respect to a subset of the states), reachability arguments can be used to drop polytopic cells that are guaranteed not to be visited during operation; this was recently studied in \cite{maddalena2019robust} and in \cite{kvasnica2019complexity}. Additional region elimination schemes were presented in \cite{rossiter2005using} and \cite{christophersen2007controller}, where an interpolation strategy and a backup control law were respectively defined to be used in case the system visited a state whose region had been eliminated. Finally, approximation methods for eMPC were investigated using polynomial functions \cite{kvasnica2011stabilizing}, multi-scale basis \cite{summers2011multiresolution}, as well as piecewise-affine functions \cite{jones2010polytopic}.

The use of Neural Networks (NN) to learn nonlinear MPC controllers was studied in \cite{parisini1995receding}, where one hidden layer and sigmoid activation functions were employed. The authors assumed a specific degree of regularity of the obtained approximator in terms of a bound of an integral weighted by its Fourier decomposition coefficients. It was then shown that the final NN approximation error can be bounded. More recently, \cite{chen2018approximating} proposed training a NN with rectified linear units by means of reinforcement learning instead of fully supervised learning. The context was that of quadratic MPC for linear systems. Their approach was shown to alleviate the training phase computations, especially since the system model can be exploited at this stage. Nevertheless, no guarantees of closed-loop stability are given or even tools to analyze it a posteriori -- a rather common drawback of machine learning approximators. 

\textit{Contribution}: Herein we propose a method to fit datapoints sampled from an MPC controller -- either in implicit or explicit form -- with a particular type of Neural \linebreak Netowrk. The architecture is defined by two linear layers and an implicit parametric quadratic program (pQP) layer, which together are able to capture any linear MPC problem with quadratic cost. In contrast with the majority of previous approaches such as \cite{rossiter2005using,christophersen2007controller,kvasnica2011clipping}, the final complexity is incremental and can be flexibly adjusted. The resulting approximator can be deployed essentially as a new simpler eMPC controller, and \textit{closed-loop stability can be certified a posteriori} with tools already available in the literature. A case study in the power electronics domain, where several successful implementations of eMPC have been reported (see e.g. \cite{mariethoz2008explicit}), is presented to illustrate the potential of the proposed methodology. Although the proposed simplification technique gives up the original controller optimality, our numerical investigations show that significant complexity reduction is possible with little to negligible performance degradation\footnote{All MATLAB scripts and \texttt{python} code can be downloaded from \texttt{www.c4science.ch/source/QPFit}.}.

\textit{Notation}: $\mathbb{R}^{n}$ is the n-dimensional Euclidean space equipped with its usual norm. Given a matrix $A$, its transpose is denoted by $A'$, and $A \succ (\succeq) \; 0$ means it is positive-definite (semidefinite). Furthermore, $||v||^2_{A}$ is defined as $||v'Av||^2$, $I$ denotes an identity matrix of appropriate size, and $\bm{0}$ denotes a zero matrix of appropriate size. $\text{diag}(a_1,\dots,a_n)$ represents a diagonal matrix with entries $a_1,\dots,a_n$.

\section{Learning Predictive Controllers}

Consider the following standard MPC formulation for linear dynamical systems
\begin{subequations}
\label{eq:mpcFormulation}
\begin{align}
    \mathds{P}1: \min_{X,U} \quad & \sum_{k=0}^{H-1} \big( x_k' Q x_k + u_k' R u_k \big) + x_H' P x_H\\
    \text{s.t.} \quad & \forall k = 0,\dots,H-1 \\
    & x_{k+1} = A x_{k} + B u_{k} \\
    & x_{k} \in \mathbb{X} \\
    & u_{k} \in \mathbb{U} \\
    & x_{H} \in \mathbb{X}_H \\
    & x_0 = x(0)
\end{align}
\end{subequations}
where $X:=\{x_1,\dots,x_{H}\}$, $U:=\{u_0,\dots,u_{H-1}\}$, $Q \succeq 0$, $P \succeq 0$, $R \succ 0$, and the constraints are all described by affine equalities and inequalities. Denote by $\bm{\pi}: \mathcal{X} \rightarrow \mathcal{U}$ the optimal solution of \eqref{eq:mpcFormulation} in its parametric form with respect to the initial conditions $x(0)$, where $\mathcal{X} \subseteq \mathbb{R}^n$ is the feasible state space of $\mathds{P}1$ and $\mathcal{U} \subseteq \mathbb{R}^m$ is the control space. We assume a set of $N$ samples can be acquired from the original control law\footnote{The dataset can also be directly obtained from the implicit controller; depending on the size of the problem at hand, computing the explicit solution might be intractable.}
\begin{subequations}
\begin{align}
    & D = \{\text{x}_i,\text{u}_i\}_{i=1}^{N} \\[3pt]
    & \text{u}_i = \bm{\pi}(\text{x}_i), \; i=1,\dots,N
\end{align}
\end{subequations}
The process of acquiring the training dataset does not have to follow a particular distribution, nor do the samples have to be independent.

\subsection{The proposed architecture}

\begin{figure}[t]
\begin{center}
    \includegraphics[width=7cm]{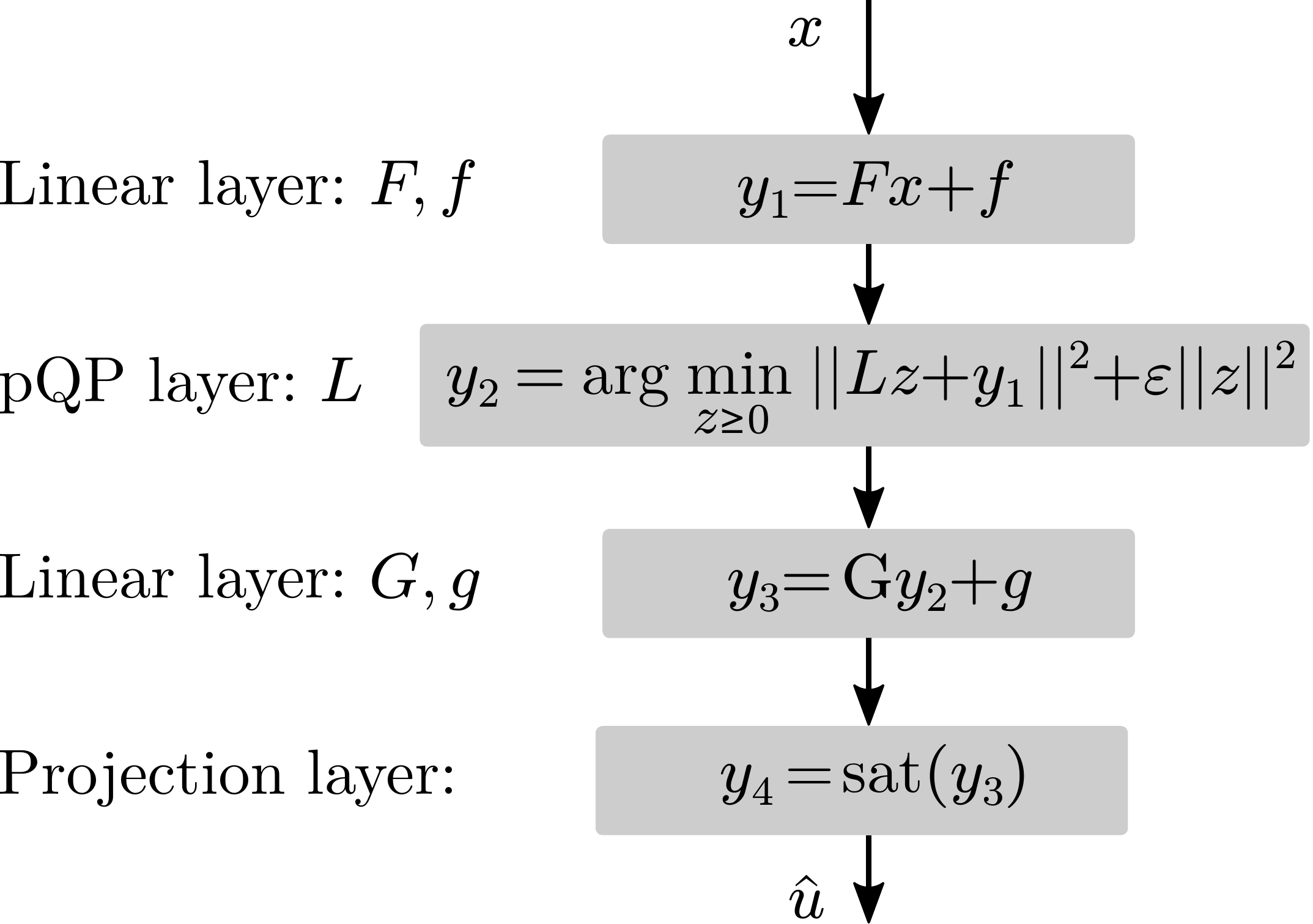}    
    \caption{An illustration of the NN eMPC controller approximator. The parameters to be optimized are shown on the left-hand side.}
    \label{fig:nnArchitecture}
\end{center}
\end{figure}

The key idea behind the proposed approximator is the use of a parametric quadratic program layer as part of the Neural Network, and optimizing over its parameters in order to fit the available dataset. This layer is implicitly described by the quadratic program
\begin{equation}
    \label{eq:pQP}
    z^\star = \text{arg }\min_{z \geq 0} \; ||Lz + y_1(x)||^{2} + \epsilon ||z||^2
\end{equation}
which is \textit{always feasible} and \textit{bounded from below}. The size of this mathematical program, i.e. the dimension of $z$, can be tuned to attain approximations with different complexity. Moreover, as notation suggests, the parameter $y_1$ depends on a previous affine layer that maps the system states into the $z$ space, $y_1 := Fx + f$. Let $y_2 := z^{\star}$, then another affine layer maps the optimal solution to the input space $y_3 := Gy_2 + g$, and a projection onto the feasible input set produces the final control action $\hat{u} := \text{Proj}_{\, \mathbb{U}}(y_2)$. This last step is necessary to guarantee feasibility of the control moves (see e.g. \cite{chen2018approximating}). An illustration of the proposed architecture is presented in Fig.~\ref{fig:nnArchitecture}, where the projection layer was particularized to a familiar element-wise saturation operation $\text{sat}(\cdot)$, valid for box input constraints.

Let the chosen number of decision variables in the pQP be $z \in \mathbb{R}^{n_z}$. We choose $L \in \mathbb{R}^{n_z \times n_z}$ to be square, and therefore $F \in \mathbb{R}^{n_z \times n}$, $f \in \mathbb{R}^{n_z}$. Moreover, $G \in \mathbb{R}^{m \times n_z}$ and $g \in \mathbb{R}^{m}$. If $n_z \ge n$, the first layer lifts the input data into a higher dimensional space before it is passed through the optimization layer. A last affine function then projects it onto the control space. These facts will be later employed to analyze the representative power of the network.

The parameters to be trained are therefore $F$, $f$, $L$, $G$ and $g$. This process can be carried out via a stochastic gradient descent algorithm applied to an appropriate loss function. Differentiability of all layers is trivial with the exception of the pQP one \cite{gould2016differentiating,amos2017optnet}. Regarding the later, note that the objective in \eqref{eq:pQP} can be rewritten as
\begin{equation}
    \label{eq:idk}
    V(z) := z'(\epsilon I + L'L)z + (2L'y_1(x))' \, z + y_1(x)'y_1(x)
\end{equation}
whose Lagrangian is simply
\begin{equation}
    \mathcal{L}(z,\lambda) = V(z) - \lambda'z
\end{equation}
The Karush-Kuhn-Tucker (KKT) conditions for primal and dual feasibility, complementary slackness, and stationarity then read 
\begin{subequations}
    \label{eq:kkt}
    \begin{align}
        & z^\star \geq 0 \\
        & \lambda^\star \geq 0 \\
        & \lambda_i^\star z_i^\star = 0, \ \forall i = 1,\dots,n_z \\
        & 2(\epsilon I + L'L)z^\star + (2L'y_1(x)) - \lambda^\star = 0
    \end{align}
\end{subequations}
where $\lambda_i$ and $z_i$ denote the components of the Lagrange multipliers and decision variables vectors. The following proposition presents the differentiability properties of the pQP layer, and holds since \eqref{eq:pQP} is a particular instance of the \texttt{OptNet} layer \cite{amos2017optnet} with strictly convex objective function.

\medbreak
\textbf{Proposition 1}:
Let $\theta := (L,y_1)$. The parametric solution $z^\star(\theta)$ of \eqref{eq:pQP} is subdifferentiable everywhere in its domain, i.e., $\partial z^\star(\theta) \neq \{ \, \}$, and $\partial z^\star(\theta)$ has a unique element (the jacobian) everywhere but in a set of measure zero.
\medbreak

As shown in \cite{amos2017optnet}, the relevant gradients with respect to the parameters to be trained can be obtained from the KKT set of equations \eqref{eq:kkt}. Hence, backward passes are possible and backpropagation can be performed to optimize all of the NN parameters.

\subsection{Properties of the approximator}

The authors of \cite{hempel2013every} showed that any continuous PWA function can be obtained as the solution of a particular parametric linear program (pLP) transformed by a linear map. Even though this view could be adopted herein, we instead prove a different result that is enough in the context of linear eMPC.

\medbreak
\textbf{Theorem 1}:
(The proposed NN architecture can learn any linear quadratic MPC controller) Let $\hat{\bm{\pi}}: \mathcal{X} \rightarrow \mathcal{U}$ be the map defined by the composition of all four layers, i.e., $\hat{\bm{\pi}}(x) := y_4 \circ y_3 \circ y_2 \circ y_1(x)$. Set $\epsilon = 0$, then $\exists F$, $f$, $L$, $G$ and $g$ with appropriate dimensions such that $\forall x \in \mathcal{X}, \ \hat{\bm{\pi}}(x) = \bm{\pi}(x)$.
\medbreak

\textit{Proof}: Define the following parametric problem
\begin{subequations}
\begin{align}
    \mathds{P}2: \min_{U} \quad & U' \, \Lambda \, U + x(0)' \, \Gamma \, U\\
    \text{s.t.} \quad & \Phi \, U \leq  \Omega \, x(0) + \omega \label{eq:constr}
\end{align}
\end{subequations}
obtained from $\mathds{P}1$ by using the equality constraints to eliminate all state decision variables (see \cite{wright2019efficient} for the construction of each matrix and vector). We also have that $\Lambda \succ 0$. Problems $\mathds{P}2$ and $\mathds{P}1$ are then equivalent in the sense that the solution $U^\star$ of $\mathds{P}2$ and $\{X^\star,U^\star\}$ of $\mathds{P}1$ have the same $U^\star$ component. Next, the dual problem of $\mathds{P}2$ can be shown to be
\begin{equation}
\begin{aligned}
    \label{eq:dualMPC}
    \mathds{D}2: \min_{\lambda \geq 0} \ \frac{1}{4} \, & \big[ \lambda'\Phi \, \Lambda^{-1}\Phi'\lambda + \, (4 x(0)'\Omega' +2 x(0)'\Gamma \Lambda^{-1} \, \Phi' \\
    & \cdots + 4 \omega')\lambda + \, x(0)'\Gamma \Lambda^{-1} \Gamma'x(0) \big]
\end{aligned}
\end{equation}
From the stationarity optimality condition of $\mathds{P}2$, it is possible to recover the primal optimal solution $U^\star$ from the dual optimal solution $\lambda^\star$
\begin{equation}
    \label{eq:stationarity}
    U^\star = -0.5 \, \Lambda^{-1}\Phi' \lambda^{\star} -0.5 \, \Lambda^{-1} \Gamma'x(0)
\end{equation}

It would be possible to enforce the first two layers to match the dual problem $\mathds{D}2$, and the third to implement \eqref{eq:stationarity} and recover the primal solution. Nevertheless, this would require the third layer to have a so called skip connection directly from the NN input, i.e., access to $x(0)$. We instead define slightly different linear and pQP layers that not only learn $\mathds{D}2$, but also let $x(0)$ pass through them and arrive at the second linear layer.

Let the auxiliary variable $\tilde{L}$ and function $\tilde{y}_1(x) := \tilde{F}x + \tilde{f}$ be the solution to (compare \eqref{eq:idk} and \eqref{eq:dualMPC})
\begin{subequations}
\label{eq:Landg1}
\begin{align}
\tilde{L}'\tilde{L} + \epsilon I & = 0.25 \, \Phi \Lambda^{-1} \Phi' \\
2 \tilde{L}'\tilde{y}_1(x) & = \Omega \, x(0) + 0.5 \, \Phi \Lambda^{-1}x(0) + \omega 
\end{align}
\end{subequations}
which leads to $\epsilon = 0$, $\tilde{L} = 0.5 \, (\Phi \tilde{\Lambda})'$, where $\tilde{\Lambda}$ is the unique square root of $\Lambda^{-1}$, guaranteed to exist as $\Lambda^{-1} \succ 0$. Then, $\tilde{y}_1(x) = (\Phi \tilde{\Lambda})^{-1} (\Omega \, x(0) + 0.5 \, \Phi \Lambda^{-1} x(0) + \omega) \implies \tilde{F} = (\Phi\tilde{\Lambda})^{-1}(\Omega+0.5 \, \Phi \Lambda^{-1}), \; \tilde{f} = (\Phi\tilde{\Lambda})^{-1}\omega$. 

Set the first layer weights to $F = [-I \ I \ \tilde{F}]'$ and $f = [\bm{0} \ \bm{0} \ \tilde{f}]'$ so that $y_1(x) = [-x; \ x; \ \tilde{F}x + \tilde{f}]$. Set the weights of the pQP layer \eqref{eq:pQP} to $\epsilon = 0$ and $L = [I \ \bm{0} \ \bm{0}; \, \bm{0} \ I \ \bm{0}; \, \bm{0} \ \bm{0} \ \tilde{L}]$. If we partition the decision variable as $z = [\tilde{z} \ x^p \ x^n]'$, this results in 
\begin{equation}
    \label{eq:pQPtilde}
    \min_{\tilde{z},x^p,x^n \geq 0} \, ||x^p - x(0)||^{2} + ||x^n + x(0)||^{2} + ||\tilde{L}\tilde{z} + \tilde{y}_1(x)||^{2} 
\end{equation}
which is a separable objective in $\tilde{z}$, $\tilde{x}^p$ and $\tilde{x}^n$. Due to the choice of $\tilde{L}$ and $\tilde{y}_1(x)$ in \eqref{eq:Landg1}, the last term of the pQP matches the dual $\mathds{D}2$ with the exception of its constant term -- not relevant for determining the optimal solution. Therefore, we have that $\tilde{z}^\star$ in \eqref{eq:pQPtilde} matches $\lambda^\star$ in $\mathds{D}2$. Regarding $x^{p\star}$, the $n$ optimizer components will satisfy $\forall i = 1,\dots,n$, $x^{p\star}_i = x_i(0)$ if $x_i(0) \geq 0$, else $x^{p\star}_i = 0$. Similarly, $x^{n\star}_i = -x_i(0)$ if $x_i(0) \leq 0$, else $x^{n\star}_i = 0$. Therefore, $x^{p\star} - x^{n\star} = x(0)$.

Next set the weights of the second linear layer $y_3$ to $G = [- 0.5 \Lambda^{-1}\Phi' \ -0.5 \Lambda^{-1}\Gamma' \ \ 0.5\Lambda^{-1}\Gamma']$ and $g = \bm{0}$. Therefore, $y_3 = G \, [\tilde{z}^\star \ x^{p \star} \ x^{n\star}]' = G \, [\lambda^\star \ x^{p \star} \ x^{n\star}]' = U^\star$, where equality \eqref{eq:stationarity} was used in the last step. Finally, note that the last layer $y_4 = \text{Proj}_{\mathbb{U}}(y_3)$ will simply evaluate to $y_3$ since $y_3$ is the optimal primal solution $U^\star$, which satisfies the constraints \eqref{eq:constr} and necessarily belongs to $\mathbb{U}$. The theorem then follows from the fact that $x(0)$ in the above calculations can be taken to be any point $x$ in $\mathcal{X}$. $\square$ \medbreak

Exactly matching the original MPC controller would require $L$ to have the same size of $\Phi \tilde{\Lambda}$ and $\epsilon = 0$ as shown. Nevertheless, we are interested precisely in reducing the complexity of the resulting controller through employing less parameters. In this process, choosing a regularizer $\epsilon > 0$ is beneficial since it ensures that the QP is bounded during the training phase for all possible parameters.

\subsection{Stability certification}

After the NN has been trained and its weights have been defined, closed-loop stability in the sense of Lyapunov can be verified through sum-of-squares (SoS) programming. First note that the NN is a composition of linear maps and optimization problems. The central idea is that the forward pass involves calculating the optimal solution to these mathematical programs, which necessarily satisfy their respective KKT conditions, e.g. \eqref{eq:kkt}. Furthermore, these conditions are sets of polynomial inequalities in the primal-dual lifted space; hence, the control moves are defined by polynomial and linear functions. As the system being controlled is linear, the certification technique presented in \cite{korda2017stability} can be readily applied.

\subsection{Controller deployment}
\label{sec:deployment}

The final approximated control law $u = \hat{\bm{\pi}}(x)$ is evaluated by a forward pass in the NN. If the first and second linear layers are incorporated respectively into the pQP and projection ones, this can be done by solving the two (simpler) optimization problems on-line. Alternatively, two explicit PWA solutions can be calculated off-line with a bound on the number of regions dependent on the chosen size $n_z$. If the constraints $u \in \mathbb{U}$ are simply a box -- as in many practical applications -- an element-wise saturation is the closed-form solution to the projection layer $u = y_3 = \text{Proj}_{\mathbb{U}}(y_2)$, further simplifying the final controller. We illustrate the use of the proposed technique in a power electronics context in the next section.

Any discrepancy between $\hat{\bm{\pi}}(x)$ and $\bm{\pi}(x)$ in the region that contains the origin will result in steady-state errors. Several techniques can be used to overcome this issue. As discussed in \cite{parisini1995receding}, the LQR solution associated with \eqref{eq:mpcFormulation} can be stored and used whenever the system is inside the associated invariant set, where no constraints are active. Alternatively, an external disturbance estimator could be adopted to account for the optimal-approximate controller mismatch.

\section{Control of a step-down converter}

In order to enhance readability and stay consistent with the standard circuits convention, notation will be overloaded. 

\subsection{Analysis and controller design}
\label{sec:analysis}
Parallelism is a key concept to increase the efficiency and power levels of electronic converters. Still, this design choice has to be followed by proper current/voltage balancing techniques to ensure that no stage is subjected to a higher electrical stress when compared to the others. 

\begin{figure}[t]
\begin{center}
    \includegraphics[width=9cm]{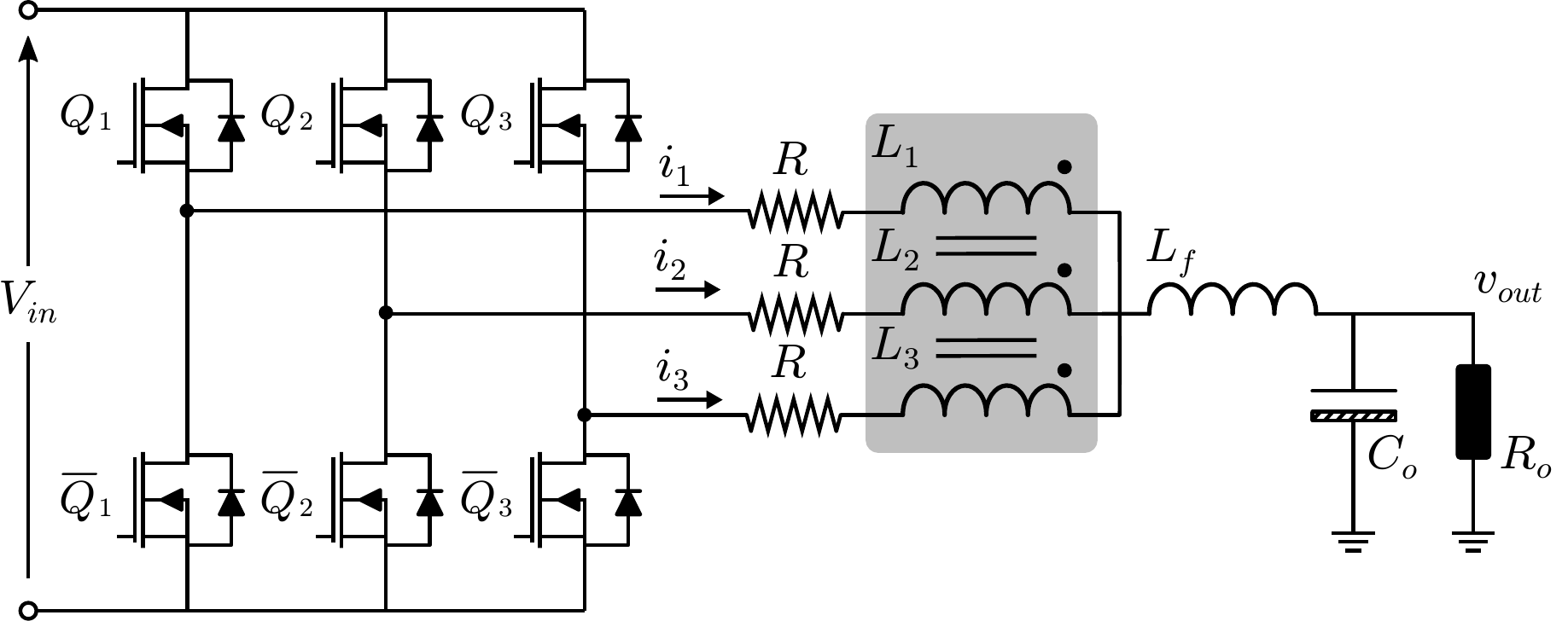}    
    \caption{Circuit diagram of the multicell step-down DC-DC converter.} 
    \label{fig:converter}
\end{center}
\end{figure}

A schematic representation of a multicell step-down converter is shown in Fig.~\ref{fig:converter}, and its parameters can be found in Table~\ref{tab:param}. The topology features three arms that are connected to a coupled inductor, and an L-C output filter. All self inductances are assumed equal $L_1 = L_2 = L_3 = L_{s}$, and all mutual inductances have value $L_m$. The switches of each arm operate in a complementary fashion at a fixed frequency $f = \,$15 kHz, and with variable but constrained duty cycle $0 \leq d_i \leq 0.9, \, i=1,2,3$. Let the average voltage applied by the arms over one switching period be denoted by $v_i := d_i V_{in}, i=1,2,3$. In order to ease the analysis, we apply the Lunze transform $\Psi$ to all variables, decomposing the phase voltages and currents into differential and common mode components
\begin{align}
    \begin{bmatrix}i_{dm1} & i_{dm2} & i_{cm}\end{bmatrix}' & := \Psi \, \begin{bmatrix}i_{1} & i_{2} & i_{3}\end{bmatrix}' \\
    \begin{bmatrix}v_{dm1} & v_{dm2} & v_{cm}\end{bmatrix}' & := \Psi \, \begin{bmatrix}v_{1} & v_{2} & v_{3}\end{bmatrix}' 
\end{align}
where $\Psi = (1/3) \, [2 \, -1 \, -1; \, -1 \ \, 2 \, -1; \, 1 \ \, 1 \ \, 1]$.

\begin{table}[t]
\begin{center}
    \caption{DC-DC converter parameters} \label{tab:param}
    \begin{tabular}{cccccccc}
        \hline 
        $V_{in}$ & $L_{s}$ & $L_{m}$ & $R$ & $L_f$ & $C_o$ & $R_o$ & \\\hline
        350$\,$V & 4$\,$mH & -2$\,$mH & 10$\,$m$\Omega$ & 270$\, \mu$H & 20$\, \mu$F & 6.25$\, \Omega$\\ \hline
    \end{tabular}
\end{center}
\end{table}

The control input is defined as $u := [v_{dm1} \; v_{dm2} \; v_{cm}]'$ and the continuous-time state vector, by appending the output voltage to the transformed currents $x := [i_{dm1} \; i_{dm2} \; i_{cm} \; v_{out}]'$. By using Kircchoff's circuit laws, a linear model of the form $\dot{x} = A_{ct} x + B_{ct} u$ can be derived with 
\smallbreak
\begin{equation}
    A_{ct} = 
    \begin{bmatrix}
        \frac{-R}{L_s + L_m} & 0 & 0 & 0 \\
        0 & \frac{-R}{L_s + L_m} & 0 & 0 \\
        0 & 0 & \frac{-R}{L_s + 2L_m + 3L_f} & \frac{-1}{L_s + 2L_m + 3L_f} \\ 
        0 & 0 & \frac{3}{C_o} & \frac{-1}{R_o C_o}
    \end{bmatrix}
\end{equation}
\begin{equation}
    B_{ct} = 
    \begin{bmatrix}
        \frac{1}{L_s} & 0 & 0\\
        0 & \frac{1}{L_s} & 0 \\
        0 & 0 & \frac{1}{L_s + 2L_m + 3L_f} \\ 
        0 & 0 & 0
    \end{bmatrix}
\end{equation}
Finally, discretization at frequency $f$ is carried out using the zero-order hold method, yielding $x_{k+1} = A x_{k} + B u_{k}$.

The control goal is to regulate the output voltage $v_{out}$ to 300$\,$V while maintaining the phase currents balanced at all times, which translates to driving the differential currents to zero. More specifically we have the following  fixed reference $x_{eq} = [0 \; 0 \; 16 \; 300]'$ with $u_{eq} = B^{\dagger}(I-A)x_{eq}$, where $B^\dagger$ is the pseudo-inverse of $B$. Moreover, the controller approximation procedure must not incur a steady-state error larger than 200$\,$mA for $i_{dm1}$ and $i_{dm2}$, and 5\% for the common mode component $i_{cm}$ and output voltage $v_{out}$. The chosen MPC cost function was
\begin{equation}
    J = \sum_{k=0}^{H-1} (||x_{k}-x_{eq}||_Q^2 + ||u_{k}-u_{eq}||_R^2) \, + \, ||x_{N}-x_{eq}||_P^2
\end{equation}
where $Q = \text{diag}(10, \, 10, \, 0.1, \, 0.1)$, $R = 0.1 \, I$, $P$ is the solution to the associated the discrete-time algebraic Riccati equation, and $H=10$. For all time instants, box state constraints were imposed $\begin{bmatrix}-5 & -5 & -10 & -20\end{bmatrix}' \leq x_{k} \leq \begin{bmatrix}5 & 5 & 30 & 400\end{bmatrix}^{\prime}$ and polyhedron constraints on the controls $H_u \, u_k \leq h_u$ that simply mapped the duty cycle saturation to the Lunze domain. Due to the polytopic input constraints, the system cannot be decomposed into three decoupled parts as the structure of matrices $A_{ct}$ and $B_{ct}$ suggest. Furthermore, the standard terminal set constraint was imposed on $x_H$, defined as the invariant set associated to the unconstrained infinite-time problem formulation.

\subsection{Learning the optimal controller}
With the aid of the Multi-Parametric Toolbox (MPT) \cite{herceg2013multi}, the optimal eMPC solution $\bm{\pi}(x)$ was calculated and consisted of 2'337 critical regions. By counting the number of parameters necessary to describe each halfspace and control gain, the memory requirement of this PWA function was found to be 518$\,$kB. In the previous calculation, a 4 byte representation was considered for both integers and floating point numbers.

\begin{figure}[t]
\begin{center}
    \includegraphics[scale=0.5]{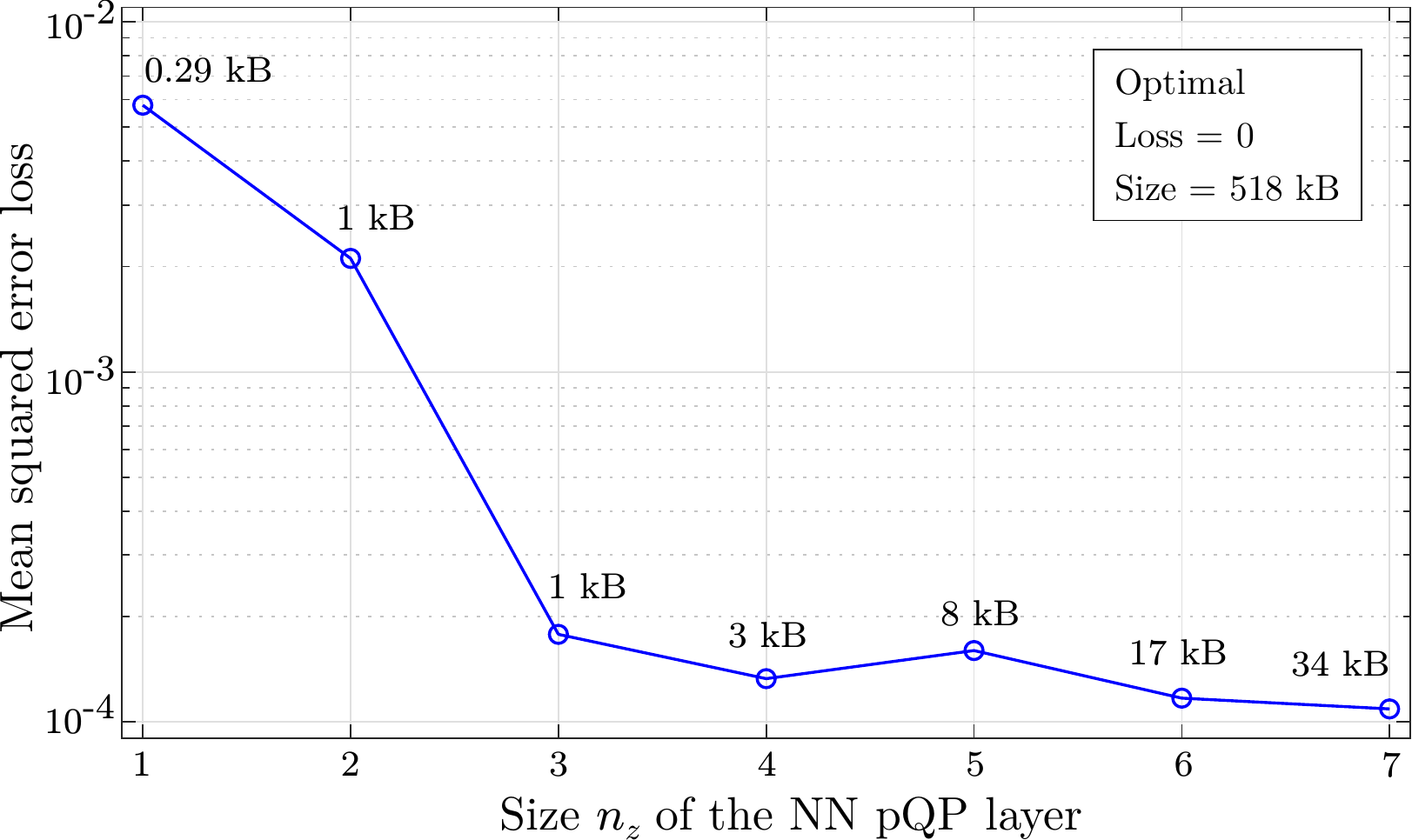}    
    \caption{Neural network training loss as a function of the pQP layer size, and storage requirements associated with their PWA representations.} 
    \label{fig:lossGraph}
\end{center}
\end{figure}

Next, 5'000 samples were randomly acquired from the eMPC controller using a uniform distribution. The first and second components of the sampled control moves had considerably smaller amplitudes compared to the third due to the structure of the Lunze transform $\Psi$. The dataset labels $\{\text{u}_i\}_{i=1}^{5000}$ had therefore to be scaled to ensure a similar learning of all control components. Moreover, instead of $\text{Proj}_{\mathbb{U}}(y_3)$, the last NN layer was simplified to $y_4 = \Psi \, \text{sat}(y_3)$ with saturation limits $0$ and $0.9 \, V_{in}$. This clearly guarantees control feasibility without the need of a second quadratic program. NN approximators were trained using \texttt{PyTorch} and the \texttt{OptNet} framework \cite{amos2017optnet}. A mean squared error loss function was minimized by employing the Adam algorithm, mini-batch stochastic gradient descent with batch size $50$, and $150$ epochs. The size $n_z$ of the pQP layer was varied from 1 to 7 and a total of 10 models were trained for each size; the lowest obtained losses are shown in Figure~\ref{fig:lossGraph}. On average, training a model required 42~minutes on a 3.1 GHz Intel Core i7 machine without GPU acceleration, and 23~minutes with a single NVIDIA Tesla T4 graphics card. The learned parameters were then exported to MATLAB in order to calculate the PWA solution of the pQP layer. 

\begin{figure}[t]
\begin{center}
    \includegraphics[scale=0.55]{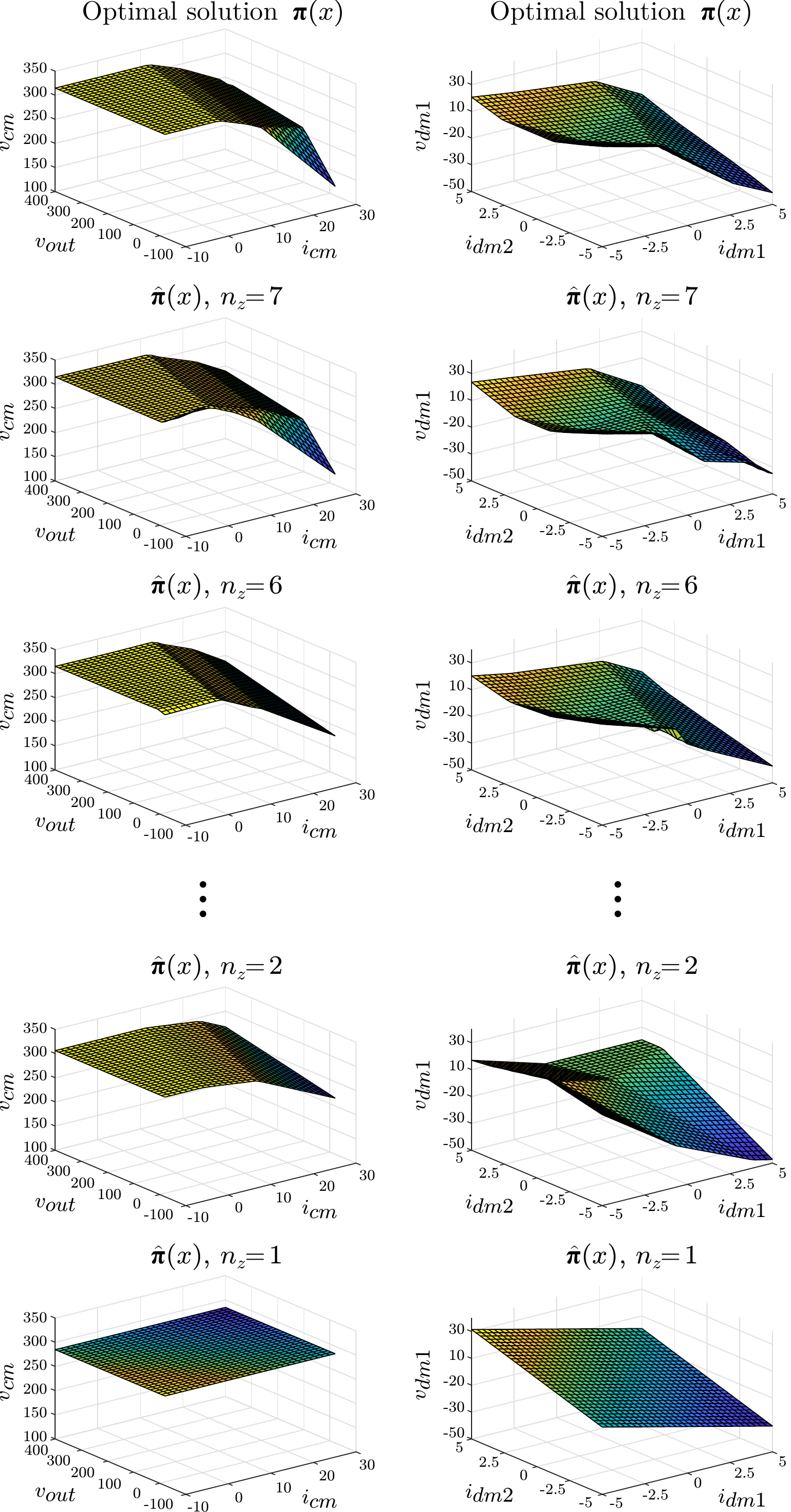}    
    \caption{Slices of the optimal eMPC controller $\bm{\pi}(x)$ and several PWA NN approximations $\hat{\bm{\pi}}(x)$. The left plots are associated to $v_{cm}$ and the right plots, to $v_{dm1}$.} 
    \label{fig:niceSlices}
\end{center}
\end{figure}

\begin{figure}[t]
\begin{center}
    \includegraphics[scale=0.6]{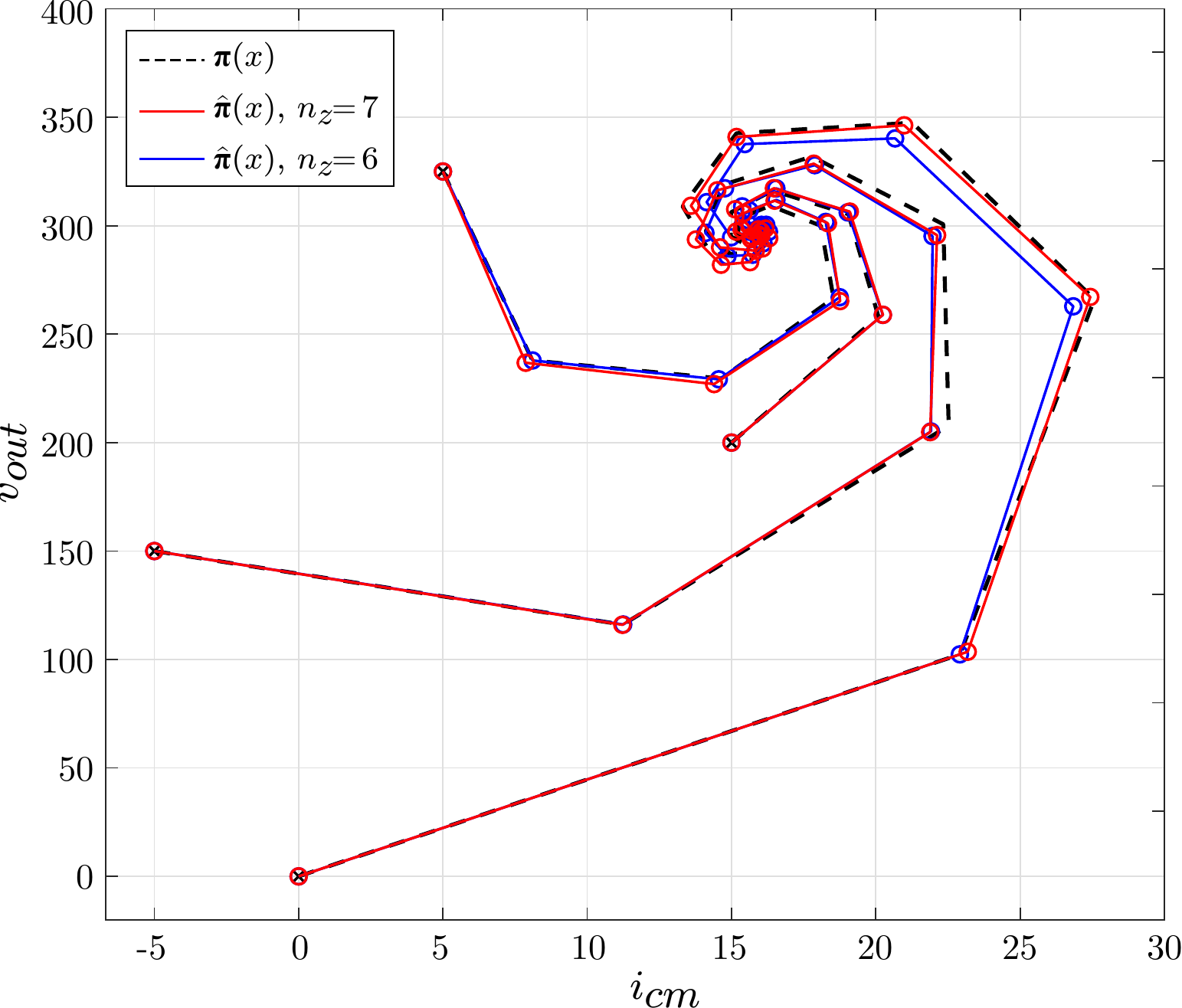}    
    \caption{Output voltage and common mode current phase portraits when employing the optimal controller and the two viable approximations.} 
    \label{fig:plot1}
\end{center}
\end{figure}

An increase in the $n_z$ size clearly expands the representation capabilities of the neural network. This however does not always translate to a decrease in the final loss since the training process is affected by the weights initialization among other factors. Although not monotonic, we see a decrease of the overall training loss in Figure~\ref{fig:lossGraph} as $n_z$ grows. In order to validate the models, the start-up response of the converter was analyzed under all 7 different approximate controllers, and only the two largest ones ($n_z=$~6 and $n_z=$~7) met the target specifications given in Section~\ref{sec:analysis}. We refer to these two solutions as the \textit{viable learned controllers}. Slices of their control surfaces are shown in Figure~\ref{fig:niceSlices}\footnote{The $v_{out}$ axis was extended to -100~V for visualization purposes.}, and a phase portrait of the closed-loop system evolution over 50 steps starting from four initial conditions is depicted in Figure~\ref{fig:plot1}. A summary of the two viable learned controller features is presented in Table~\ref{tab:sec}, including the number of polytopic regions, the storage requirements, the worst-case computation time\footnote{Before proceeding to implementation, a further speed up is possible through the methods listed in the Introduction.} and the steady state (SS) error for $i_{cm}$ and $v_{out}$ -- both clearly always equal. Even though four initial states were given, the systems always converged to the same points and, hence, only one SS error is reported. Plus, the storage numbers also take into account all the remaining layers parameters. Analyzing the obtained results we see that the approximations drastically reduced the storage requirements by 93.4\% and 96.7\% and sped up the worst-case evaluation time by 83.7\% and 88.4\%, respectively for the $n_z=$~7 and $n_z=$~6 cases. The closed-loop trajectories with the proposed $\hat{\bm{\pi}}(x)$ remained reasonably close to the scenario with the optimal $\bm{\pi}(x)$, converging to nearby equilibrium points. In practice, steady-state errors could be removed by using the tools mentioned in Section~\ref{sec:deployment}.

\begin{table}[t]
\begin{center}
    \caption{Optimal and viable learned controllers information} \label{tab:sec}
    \begin{tabular}{lcccc}
        \cline{2-5}
         & Regions & Storage & Comp. time & SS error \\\hline
        $\bm{\pi}(x)$ & 2'337 & 518$\,$kB & 12.9$\,$ms & 0\% \\ \hline
        $\hat{\bm{\pi}}(x)$, $n_z$ = 7 & 107 & 34$\,$kB & 2.1$\,$ms & 0.59\% \\ \hline
        $\hat{\bm{\pi}}(x)$, $n_z$ = 6 & 56 & 17$\,$kB & 1.5$\,$ms & 1.25\% \\ \hline
    \end{tabular}
\end{center}
\end{table}

\section{Conclusion}

We proposed a novel method to approximate explicit MPC controllers from samples of the optimal control law. The essence of the approach is a Neural Network architecture featuring a pQP layer incorporated to learn the MPC dual problem. It was shown that, with an appropriate pQP size, it is possible to learn any linear quadratic MPC exactly. After its training, an equivalent representation of the NN can be found off-line as a new and simpler controller. Steady-state errors due to approximation inaccuracies can be mitigated by employing standard tools. A numerical example was provided where the storage requirements and computational burden of the approximate PWA controllers were significantly lower than their optimal counterpart with minor performance degradation.

\bibliographystyle{IEEEtran}
\bibliography{ifacconf} 

\end{document}